\documentstyle[twoside,fleqn,espcrc2,epsf,amssymb]{article}

\newcommand{\AmS}{{\protect\the\textfont2
  A\kern-.1667em\lower.5ex\hbox{M}\kern-.125emS}}

\title{Correlations and Binding in 4D Dynamical Triangulation}

\author{Bas V.~de Bakker\address{Centre for High-Energy Astrophysics,
        Kruislaan 403, 1098 SJ Amsterdam, The Netherlands}\thanks{\tt
Affiliation before September 1995: Institute for Theoretical Physics, 
University of Amsterdam}
        and 
        Jan Smit\address{Institute for Theoretical Physics, 
        University of Amsterdam, \\ 
        Valckenierstraat 65, 1018 XE Amsterdam, 
        The Netherlands}\thanks{\tt
        Presented by J.\ Smit}}
       
\begin{document}

\begin{abstract}
We study correlations on the euclidean spacetimes generated in 
Monte Carlo simulations of the model. In the elongated phase, 
curvature correlations appear to fall off like a fractional 
power. Near the transition to the crumpled phase this power is 
consistent with 4. We also present improved data of our 
computations of the binding energy of test particles.
\end{abstract}

\maketitle

\section{Introduction}
The canonical partition function of the dynamical triangulation 
model is defined as a sum over simplicial manifolds consisting of 
equilateral four-simplices glued together according to 
triangulations ${\cal T}$,
\begin{equation}
Z = \sum_{{\cal T}(N_4)} \exp(\kappa_2 N_2).
\end{equation}
Here $N_2$ is the number of triangles in the simplicial manifold 
consisting of $N_4$ four-simplices and the topology is chosen to 
be that of $S^4$. (For more details see \cite{Catterall}.)
We recall that the system has two phases, a crumpled phase at low 
$\kappa_2$ and an elongated phase at high $\kappa_2$. 

The model is supposed to 
represent the quantum gravitational path integral over euclidean 
spacetimes weighted with the Regge-Einstein action,
with 
volume 
$V\propto N_4$ and bare Newton constant $G_0 \propto \kappa_2^{-1}$.
As such it should be able to reproduce semiclassical Einstein 
gravity as an effective theory. This can presumably be investigated
using coordinate invariant correlation functions. Another test is 
to see if scalar test particles form bound states, 
with appropriate binding energies under nonrelativistic conditions.  

In \cite{1} we reported on such preliminary binding energy 
calculations. More insight was needed at that time about the 
spacetimes described by the model. For a semiclassical 
interpretation in terms of an effective action it is obviously 
important if there is in some sense a background spacetime which 
resembles the classical $S^4$. To address this question we 
studied \cite{2} the number of simplices at geodesic distance 
$r$, $N'(r)$. We found encouraging scaling of this function in a region
around the transition between the crumpled and elongated phase. (For
a different scaling analysis see \cite{AmJu}.)
At the transition we observed that $N'(r)$ behaves approximately like a 
four-sphere, 
$\propto \sin^3[(\pi/2) r/r_m]$,
for distances $r$ outside a short distance `planckian regime'  
up to intermediate distances of order $r_m$. For larger distances 
fluctuations distort the shape of $N'(r)$.
This suggests that we focus our attention on this 
intermediate distance regime for a possible semiclassical 
interpretation of the data.

\section{Correlations}

Let $n_x$ denote the number of four-simplices
winding around a triangle $x$ (often called a `hinge').
We measured correlations \cite{3} 
\begin{equation}
C_O = \langle O\, O\rangle(r) - [\langle O\, 1 \rangle (r)]^2, \label{e1}
\end{equation}
where the brackets denote the average over simplicial 
configurations, 
\begin{equation}
\langle O\,O\rangle(r) = \langle 
\frac{\sum_{xy} O_x O_y \delta_{d_{xy}-r} }
{\sum_{xy} \delta_{d_{xy} -r} }
\rangle, \label{e2}
\end{equation}
{
\thispagestyle{myheadings}
\renewcommand{\thepage}{ITFA-95-15}

\clearpage
}
with $O_x = n_x^{-1}$, $n_x$, and similar for $\langle O\, 
1\rangle$. Here $d_{xy}$ is a `triangle geodesic distance' 
between $x$ and $y$. The `disconnected part' subtracted in 
(\ref{e1}) has the form of a correlation of the observable $O_x$ 
with the distance to a point $y$, $d_{xy}$, which is itself a 
function of the geometry. This at first sight unfamiliar form is 
needed for the correlation to vanish at larger distances. 
 
\begin{figure}
  \epsfxsize=75mm \epsffile{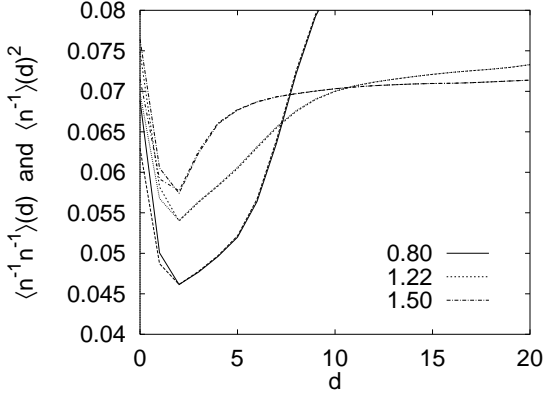}
\caption{
$\langle n^{-1}\, n^{-1}\rangle(d)$ and $\langle n^{-1}\, 1\rangle^2(d)$.
}
\label{f1}
\end{figure}

Other expressions for similar correlations are conceivable. For 
example, including a volume element $\propto n_x$ with a 
summation over $x$ may appear more natural from the Regge 
calculus point of view. We shall argue below that this leads to 
the same scaling results. Taking the average $\langle \;\rangle$ 
separately in numerator and denominator is another option, which 
should not make significant difference for not too large 
distances, because of the self-averaging effect of the summations in
(\ref{e2}).

In Regge calculus $n_x^{-1}$ is related to the scalar
curvature $R_x$, and $n_x$ to a local volume element at $x$;  we 
write therefore $C_R$ ($C_V$) for $O_x = n^{-1}_x$ ($O_x = n_x$). 
However, generally, in a scaling region, a lattice operator is 
expected to reduce to a combination of various continuum 
operators, weighted with powers of the lattice distance according 
to their dimensions. Assuming that the theory can be described by 
a continuum metric tensor $g_{\mu\nu}$ with corresponding 
curvature $R$, the dominant continuum observable corresponding to 
both our lattice correlation functions $C_R$ and $C_V$ would be 
given by
\[    \frac{
    \langle \int dx\, dy
    \sqrt{g(x)} \sqrt{g(y)} \, \delta( d(x,y) - r ) R(x) R(y)
    \rangle
    }{ \langle
    \int dx\, dy \sqrt{g(x)} \sqrt{g(y)} \, \delta( d(x,y) - r )
    \rangle
    }
\]\[
-   \left[
    \frac{
    \langle \int dx\, dy
    \sqrt{g(x)} \sqrt{g(y)} \, \delta( d(x,y) - r ) R(x)
    \rangle
    }{ \langle
    \int dx\, dy \sqrt{g(x)} \sqrt{g(y)} \, \delta( d(x,y) - r )
    \rangle
    }
    \right]^2.
\]
Here $d(x,y)$ is the geodesic distance between the points $x$ and $y$
for a given metric $g_{\mu\nu}$, and the average over geometries
is supposed to be calculable in semiclassical fashion. 
Of course, we do not know the effective action specifying this average.  
It could be a combination of $\int dx \sqrt{g} R$ and higher order 
terms like $R^2$.

\begin{figure}
  \epsfxsize=75mm \epsffile{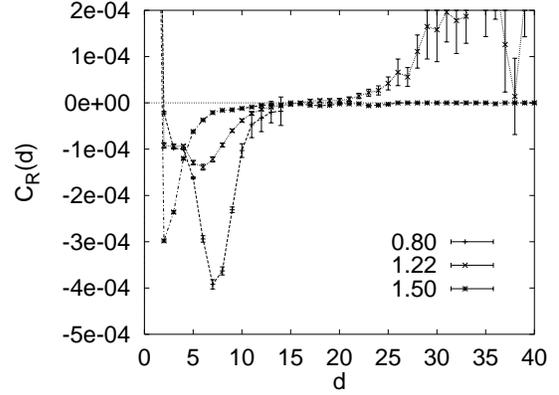}
\caption{
$C_R(d)$ corresponding to fig.\ \protect \ref{f1}.
}
\label{f2}
\end{figure}

Fig.\ \ref{f1} shows the two terms in eq.\ (\ref{e1}) separately for
$C_R(r)$ for $N_4 = 16000$ and $\kappa_2 = 0.80$ (crumpled phase),
1.22 (transition) and 1.50 (elongated phase). The two contributions are
very similar and fig.\ \ref{f2} shows their difference, $C_R(r)$, which
drops to zero before finite size effects take over.

\begin{figure}
  \epsfxsize=75mm \epsffile{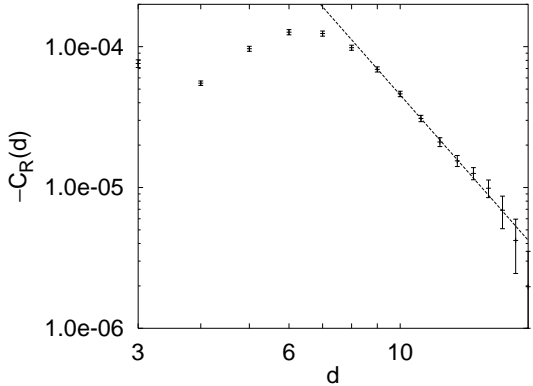}
\caption{
Power fit to $C_R(d)$ near the transition.
}
\label{f3}
\end{figure}

Fig.\ \ref{f3} shows $C_R(r)$ for a system of
$N_4 = 32000$ simplices and $\kappa_2=1.255$, which is at the 
transition between crumpled and elongated phase. For this case, with
the employed definition of geodesic distance \cite{3},
the `planckian regime' mentioned in the Introduction is 
$r\lesssim 7$, while $r_m \approx 11$. So the approximate $S^4$ 
background geometry is to be found in the region 
$7\lesssim r \lesssim 11$.

We see an 
approximate power behavior $C_V \approx a r^b$ for $r\gtrsim 7$. 
A fit in the range $8\leq r\leq 18$ gives $a = -0.5(2)$, 
$b=-4.0(2)$. The plot for $C_V$ looks similar and leads to a 
compatible power $b=-4.30(12)$, with $a=-5.7(1.6)$. The equality 
of the powers supports the argument given above that $C_R$ and 
$C_V$ are dominated by the same operators in the scaling region. 

The power law behavior of the data may be expected for a flat 
spacetime background, with $C_{V,R} \propto Gr^{-4}$ or $\propto 
G^2 r^{-8}$ for one- or two-graviton exange. The approximate 
$S^4$ interpretation of the spacetimes near the transition 
implies finite size corrections to the power law. An $S^4$ background 
curvature $\overline{R}$ would also allow a behavior $G^2 
\overline{R} r^{-4}$, by dimensional reasoning.

\begin{figure}
  \epsfxsize=75mm \epsffile{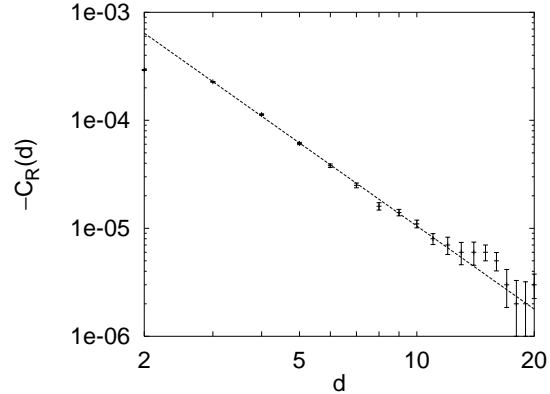}
\caption{
$C_R(d)$ in the elongated phase.
}
\label{f4}
\end{figure}

We measured $C_{V,R}$ also in the elongated phase at $\kappa_2 = 
1.50$, where the system is known to behave like a branched 
polymer with effective dimension $2$. 
As fig.\ \ref{f4} shows, a power law looks better in this phase 
and a fit in the region $3 \leq r\leq 15$ leads to $b=-2.56(3)$ 
for $C_R$ and $b=-2.57(2)$ for $C_V$. Perhaps this can be 
interpreted in an effective conformal mode theory \cite{AnMo92}.
In the crumpled phase the correlation data are difficult to 
analyse because the distances are rather short in this phase.

We draw the optimistic conclusion from the power behavior that 
the correlation functions indicate the presence of massless 
excitations.

\section{Binding}

The behaviour of test particles in the spacetimes generated with the 
dynamical triangulation model is another interesting subject.
Test particles have no backreaction on the geometry and a 
calculation of their attraction due to fluctuations is analogous to
the quenched or valence approximation in lattice QCD. Let 
$G_{xy}$ be a scalar field propagator on a given simplicial 
manifold, $(-\Box_{xy} + m_0^2 \delta_{xy})\, G_{yz} = 
\delta_{xz}$, where now $x,y$ denote the points on the dual 
lattice and $\Box$ is the corresponding lattice laplacian. We measured
\begin{eqnarray}
G(r)  &=& \langle \frac{\sum_y \delta_{d_{xy}-r}\, G_{xy} }
                            {\sum_y \delta_{d_{xy}-r}          }
                     \rangle,\nonumber\\
G^{(2)} (r) &=&\langle \left[
                            \frac{\sum_y \delta_{d_{xy}-r}\, G_{xy} }
                                 {\sum_y \delta_{d_{xy}-r}          }
               \right]^2 \rangle,
\end{eqnarray}
where $d_{xy}$ is the (dual lattice) geodesic distance between 
four-simplices. On a flat background one would expect the 
behavior $G(r) \propto r^{\alpha}\exp(-mr)$, $G^{(2)}(r) \propto 
r^{\beta} \exp(-Mr)$ at large distances, with $m$ the 
(renormalized) mass of the scalar particle and $M$ the mass of a 
two-particle state (in the quenched approximation). This is a 
bound state if $M < 2m$, with binding energy $E_b = 2m - M > 0$. 

\begin{figure}
  \epsfxsize=75mm \epsffile{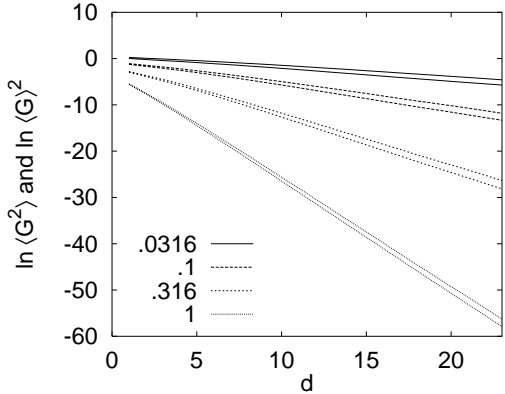}
\caption{
$G^{(2)}(d)$ and $G(d)^2$.
}
\label{f5}
\end{figure}

 Fig.\ \ref{f5} shows 
$G(r)^2$ and $G^{(2)}(r)$ for $N_4 = 32000$ at the transition, 
$\kappa_2=1.255$, for several values of the bare mass,
$m_0 = 1$, 0.316, 0.1, 0.0316. The approximate $S^4$ behavior mentioned
in the Introduction is now found in the region
$10 \lesssim r\lesssim 21$, with the distance
definition emplyed here. 
We see roughly exponential behavior suggesting that the $S^4$ 
character of the background geometry can in a first approximation 
be neglected, similar to the correlation functions above. The 
corresponding renormalized masses are $m\approx 1.19$, 0.57, 
0.27, 0.14. The ratio $m/m_0$ appears to increase substantially
as $m_0$ gets smaller. 

If the model describes semiclassical gravity we may expect 
hydrogen atom like behavior in a nonrelativistic situation, with 
small fine structure constant $\alpha\rightarrow Gm^2$: 
\begin{equation}
E_b = \frac{1}{4}\, G^2 m^5. \label{e3}
\end{equation}
In this way we may be able to define the renormalized Newton constant $G$. 
To get a rudimentary feeling for the corrections to (\ref{e3})  the hamiltonian
$H=2\sqrt{m^2 + p^2} - Gm^2/r$ may be useful. 
Replacing $p\rightarrow 1/r$ and minimizing
the energy leads to $E_b = 2m-2m\sqrt{1-G^2 m^4/4}$, which suggests that
$Gm^2 = 2$ has to be considered `large'.

\begin{figure}
  \epsfxsize=75mm \epsffile{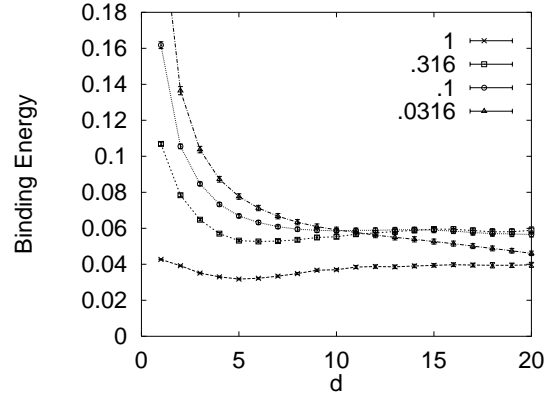}
\caption{
Effective binding energies.
}
\label{f6}
\end{figure}

Fig.\ \ref{f6} shows the effective binding energy
\begin{equation}
E_b^{\rm eff} = \frac{1}{r}\, \ln \frac{G(r)^2}{G^{(2)}(r)}.
\end{equation}
Clearly, there is no indication of nonrelativistic $m^5$ 
behavior. 
To judge the situation we have to make sure that $Gm^2$ is not too 
large (assuming $G$ exists in the model), or $m$ too small for 
neglecting finite size $S^4$ corrections, or both. It is also
desirable to get a semiclassical understanding of the quenched approximation. 
Finally, we may have to take into consideration that the euclidean 
effective action at intermediate distances
(neglecting higher order terms like $R^2$) 
is unbounded.

{\bf Acknowledgements}
We thank P.~Bia{\l}as 
for useful discussions. The simulations were carried
out on the Parsytec PowerXplorer at IC3A and the IBM SP1 at SARA.
This work is supported
in part by FOM.


\begin{thebibliography}{9}
\bibitem{Catterall} S.~Catterall, these proceedings.
\bibitem{1} B.V.~de Bakker and J.~Smit, Nucl. Phys. B. (Proc. 
Suppl.) 34 (1994) 739.
\bibitem{2}  B.V.~de Bakker and J.~Smit, Nucl. Phys. B439 (1995) 239.
\bibitem{AmJu} J.~Ambj{\o}rn and J.~Jurkiewicz, {\em Scaling in four
dimensional quantum gravity}, NBI-HE-95-05.
\bibitem{3} B.V.~de Bakker and J.~Smit, {\em Two-point
functions in 4D dynamical triangulation}, ITFA-95-1, 
to be published in Nucl. Phys. B.
\bibitem{AnMo92} I.~Antoniadis and E.~Mottola, Phys. Rev. D45 (1992) 2013.

\end{thebibliography}
\end{document}